\documentstyle[12pt,titlepage]{article}

\font\grande=cmr9.5 scaled \magstep4
\font\medio=cmr9.5 scaled \magstep2
\outer\def\beginsection#1\par{\medbreak\bigskip
      \message{#1}\leftline{\bf#1}\nobreak\medskip
\vskip-\parskip
      \noindent}

\def\laq{\raise 0.4ex\hbox{$<$}\kern -0.8em\lower 0.62
ex\hbox{$\sim$}}
\def\gaq{\raise 0.4ex\hbox{$>$}\kern -0.7em\lower 0.62
ex\hbox{$\sim$}}

\begin{document}
\bibliographystyle {unsrt}

\titlepage

\begin{flushright}
CERN-PH-TH/2004-185
\end{flushright}

\vspace{15mm}
\begin{center}
{\grande Magnetized birefringence and CMB polarization}\\
\vspace{15mm}
 Massimo Giovannini \\
\vspace{6mm}

{\sl Department of Physics, Theory Division, CERN, 1211 Geneva 23, Switzerland}

\end{center}

\vskip 2cm
\centerline{\medio  Abstract}

\noindent
The polarization plane of the cosmic microwave background radiation  
can be rotated either in a magnetized plasma 
or in the presence of  a quintessential background with pseudoscalar
coupling to electromagnetism. 
A unified treatment of these two phenomena is presented for 
cold and warm electron-ion plasmas at the pre-recombination epoch. 
The electron temperature is only relevant 
to the relativistic correction of the cold plasma results. 
The spectrum of plasma excitations is obtained from a generalized 
Appleton--Hartree equation, describing simultaneously 
the high-frequency propagation of electromagnetic waves 
in a magnetized plasma with a  dynamical quintessence field. It is 
shown that these two effects are comparable for the plausible
range of parameters allowed by present constraints. It is then argued that
the generalized expressions derived in the present study may be relevant
for direct searches of a possible rotation of the  cosmic microwave background 
polarization.
\vspace{5mm}

\vfill
\newpage
If  linearly polarized radiation passes through a cold (or warm) plasma 
containing a magnetic field, the polarization plane of the 
wave can be rotated since the two circular polarizations (forming the linearly 
polarized beam) 
are travelling at different speeds. This effect has been studied in a variety 
of different (but related) frameworks even in relativistic QED plasmas 
and in the presence of extra-dimensions (see, for instance, \cite{f1}).

The cosmic microwave background (CMB) has a weak degree of linear 
polarization, which is the direct  result of Thompson scattering. 
If CMB is linearly polarized, then its polarization plane 
can be rotated provided a sufficiently strong magnetic field 
is present around the time of decoupling \cite{f2}. Typical 
values of magnetic fields, compatible with other astrophysical 
constraints, are of the order of $B_{0}\,\,\,\laq\,\,\, 10^{-4}\,\,\,{\rm G}$ 
at the decoupling time (see, for instance, \cite{f3}).

An apparently unrelated possibility is that the quintessence 
field $\sigma$ has pseudoscalar couplings to electromagnetism \cite{q1}.
Couplings to ordinary matter, even if suppressed by the Planck scale, may lead 
to observable long-range forces and time dependence of the constants 
of nature. An approximate global symmetry may suppress these couplings 
even further as argued in \cite{q2} (second reference).
This possibility may also imply a rotation 
of the plane of polarized emission, and radio-astronomical 
implications of this type of cosmic birefringence were investigated 
through various steps \cite{q2}. Typical values of the mass of $\sigma$ are 
of the order of $10^{-33}$ eV in such a way that, between redshifts 
$0$ and $3$, $\sigma$ can start dominating the background with
energy density $m^2 \sigma_0^2 \sim \Lambda^4$, where 
$ \Lambda \simeq 10^{-3}\,\,\,
{\rm eV}$ and $ \sigma_{0} \simeq M_{\rm P} \sim 10^{18} \,\,\,{\rm GeV}$.
The effect of parity violating interactions on CMB polarization has been 
recently discused in \cite{easson}.

In this paper a unified discussion of Faraday rotation and cosmic 
birefringence will 
be presented for the specific case of the pre-recombination 
plasma. The possible interference of these two effects 
will be scrutinized in both the cold and warm plasma 
approximations. A related aspect of the present analysis will be to 
study the range of validity of the Faraday rotation estimates.

Let us start by discussing the typical scales involved in the problem.
Right before decoupling, the temperature of the plasma 
is of the order of $0.3$ eV and the typical free 
electron density can be estimated as 
\begin{equation} 
n_{\em e} \simeq x_{\rm e} ( \Omega_{{\rm b}} h_{0}^2) ( 1 + z)^3 \times 10^{-5}
\,\,\,\, {\rm cm}^{-3},
\label{eldens}
\end{equation}
where $x_{e}$ is the ionization fraction, $h_{0}$ is the indetermination 
associated with the Hubble constant, $\Omega_{b}$ is the fraction 
of critical density in baryons, and $z$ is the redshift. For typical values
of the parameters the electron density 
is of the order of $n_{\rm e} \sim 10^{3}\,\,\,\,\, {\rm cm}^{-3}$. 

Right before decoupling, the plasma is globally neutral and the ion density 
equals the electron density, i.e. $n_{\rm i} \simeq n_{\rm e} = n_{0}$, 
denoting with $n_{0}$ the common electron--ion number density.  
The global neutrality of the plasma occurs for typical length scales 
$ L \gg \lambda_{\rm D}$ where
\begin{equation}
\lambda_{\rm D} = \sqrt{ \frac{T_{\rm ei}}{8\pi e^2 n_{0}}} \simeq 
 10\,\, \biggl(\frac{n_{0}}{10^{3}\,\,\, {\rm cm}^{-3}}\biggl)^{-1/2} 
\biggl( \frac{T_{\rm ei}}{0.3 \,\,\,{\rm eV}}\biggr)^{1/2} \,\, {\rm cm}.
\label{debye}
\end{equation}
is the Debye screening length whose value is much smaller, for the same 
choice of physical parameters, than the 
electron and photon mean free paths, i.e. $\ell_{\rm e} \simeq
 5.7 \times 10^{7}\,\,\,{\rm cm}$ 
 and 
$\ell_{\gamma} \simeq 10^{4} (1+z)^{-2} (\Omega_{\rm b}h_{0}^2)^{-1}\,\,{\rm Mpc}$.

If the plasma is not magnetized,
 the only relevant frequency scales of the problem are the plasma 
frequencies, which can be 
constructed from the electron and ion densities, i.e. 
\begin{equation}
 \omega_{\rm pe} = \sqrt{\frac{4 \pi e^2 n_0}{m_{\rm e}}} \simeq
2\,\, \biggl( \frac{n_{e}}{10^{3}\,\, {\rm cm }^{-3}} \biggr)^{1/2} \,\, {\rm MHz},
\,\,\,\,\,\,\,\,\,\, \omega_{\rm pi}= 
\sqrt{ \frac{4 \pi e^2 n_0}{m_{\rm i}}} 
\simeq 40\,\, \biggl( \frac{n_{e}}{10^{3}\,\, {\rm cm }^{-3}} 
\biggr)^{1/2} \,\, {\rm kHz},
\label{plei}
\end{equation}
where, for practical estimates, the numerical value
 $ m_{\rm i} \simeq m_{\rm p} =
 0.94\,\,{\rm GeV}$ has been assumed.

The  frequencies given in Eq.
 (\ref{plei}) enter the dispersion relations, determining
the group velocity of an electromagnetic signal in the  plasma. The 
plasma frequencies for both electrons and ions are much larger than the 
collision frequencies constructed from the inverse of the mean free paths, 
i.e. $\omega_{\rm ce} \sim 520\,\,{\rm Hz}$ and $\omega_{\rm ci} \sim 2.4\,\,
{\rm Hz}$. The plasma can then be described, to a very 
good approximation, within a two-fluid framework \cite{pl1,pl2}.

If the plasma is magnetized, two new frequency scales arise 
in the problem, namely the electron and ion gyrofrequencies, i.e.
\begin{equation}
 \omega_{\rm Be} =\frac{e B_{0}}{m_{\rm e} c} \simeq 
18.08 \biggl(\frac{B_0}{ 10^{-3}\,\,\, {\rm G}}\biggr)\,\,\,{\rm kHz},
\,\,\,\,\,\,\,\,\,\,\,  \omega_{\rm Bi}=\frac{e B_{0}}{m_{\rm i} c}  
\simeq 9.66  
\biggl(\frac{B_0}{10^{-3}\,\,\, {\rm G}}\biggr)\,\,\,{\rm Hz},
\label{Bei}
\end{equation}
where $B_{0}$ is the magnetic field strength at 
the corresponding epoch.

The electron and ion gyrofrequencies, together with the plasma 
frequencies of Eq. (\ref{plei}), affect
the dispersion relations in the case of a magnetized plasma.
The maximum of the CMB can be determined, from the Wien law, around 
$1.7$ GHz; we will then be interested in the frequency range between, say, $1$
and $100$ GHz. In this frequency range $ \omega_{\rm CMB} \gg \omega_{\rm pe} >
\omega_{\rm Be}$. Since $\omega_{\rm pe} > \omega_{\rm pi}$ and 
$\omega_{\rm Be} > \omega_{\rm Bi}$, we will also have 
$\omega_{\rm CMB} \gg \omega_{\rm pi} > \omega_{\rm Bi}$ for the fiducial
set of parameters introduced so far.

Defining the appropriately
rescaled electron and ion densities, $n_{\rm e} = a^{3} \tilde{n}_{\rm e}$ 
and $n_{\rm i} = a^{3} \tilde{n}_{\rm i}$ in 
a conformally flat geometry of Friedmann--Robertson--Walker background 
characterized by a scale factor $a(\eta)$ and by the line element
$ds^2 = a^2(\eta)[ d\eta^2 - d\vec{x}^2]$,
the continuity equations for the charge densities reads 
\begin{eqnarray}
&& n_{\rm e}' + 3 w_{\rm e} {\cal H} n_{\rm e} 
+ ( w_{\rm e} + 1 ) \vec{\nabla}\cdot( n_{\rm e} \vec{v}_{\rm e} ) =0,
\label{ne}\\
&& n_{\rm i}' + 3 w_{\rm i} {\cal H} n_{\rm i} 
+ ( w_{\rm i} + 1 ) \vec{\nabla}\cdot( n_{\rm i} \vec{v}_{\rm i} ) =0,
\label{np}
\end{eqnarray}
where ${\cal H} = a'/a$; the prime denotes a derivation 
with respect to the comformal time coordinate $\eta$;
$w$ is the  barotropic index for the electron or ion fluid.

Both electrons and ions are non-relativistic before decoupling. 
Hence the barotropic index $w$ will 
be close to zero to a good approximation. 
For instance, the energy and pressure densities 
of an ideal electronic gas are given by 
\begin{equation}
\rho_{\rm e} = n_{\rm e} \biggl( m_{\rm e} 
+ \frac{3}{2} T_{\rm e}\biggr),\,\,\,\,\,\,\,\,\,\,
p_{\rm e} = n_{\rm e} T_{\rm e},
\end{equation}
and since $ w_{\rm e,i} = T_{\rm e,i}/m_{\rm e,i}$, $ w_{\rm e,i} \ll 1$ as far 
as $ T_{\rm e,i} \ll m_{\rm ,ie}$. 
 
In the cold plasma approximation the temperature 
of the ions and of the electrons vanishes. In the warm  plasma 
approximation the temperature of the two charged species may be 
very small but non-vanishing. The warm plasma 
 treatment will lead,  in practice, only 
to an effective correction of the plasma frequency. 
Since the cold plasma results turn out to be, a posteriori, rather 
accurate, the discussion will be presented in terms of the 
cold plasma description, while the results of the warm plasma 
treatment will be stated later.

To have a self-consistent set of two-fluid equations, Eqs. 
(\ref{ne}) and (\ref{np}) will be supplemented by the evolution 
equations of the velocity fields and  of 
the electromagnetic field, namely
\begin{eqnarray}
&& \rho_{\rm e} [ \vec{v}_{\rm e}' + {\cal H} \vec{v}_{\rm e} 
+ (\vec{v}_{\rm e}\cdot\vec{\nabla})\vec{v}_{\rm e} ] = - n_{\rm e}
e \biggl( \vec{E} + \frac{\vec{v}_{\rm e}}{c} \times \vec{B}\biggr),
\label{ve}\\
&&   \rho_{\rm i} [ \vec{v}_{\rm i}' + {\cal H} \vec{v}_{\rm i} 
+ (\vec{v}_{\rm i}\cdot\vec{\nabla})\vec{v}_{\rm i} ] =  n_{\rm i}
e \biggl( \vec{E} + \frac{\vec{v}_{\rm i}}{c} \times \vec{B}\biggr),
\label{vi}
\end{eqnarray}
where $\vec{E} = a^2 \vec{{\cal E}}$  and $ \vec{B} = a^2 \vec{{\cal B}}$
are the conformally rescaled electromagnetic fields obeying the following 
set of generalized Maxwell equations:
\begin{eqnarray}
&& \vec{\nabla} \cdot \vec{E} = 4 \pi e ( n_{\rm i} - n_{\rm e} ) 
+ \frac{\beta}{M} \vec{\nabla}\sigma \cdot \vec{B},
\label{div1}\\
&&\vec{\nabla} \cdot \vec{B} =0,\,\,\,\,\,\,\,\,\,\,\,\vec{\nabla} \times \vec{E} = 
- \frac{1}{c} \vec{B}'
\label{bianchi1}\\
&& \vec{\nabla} \times  \vec{B} = \frac{1}{c} \vec{E}' - \frac{\beta}{M} 
\biggl[\sigma' \vec{B} + \vec{\nabla}\sigma \times \vec{E} \biggr] 
 + \frac{ 4 \pi e}{c} ( n_{\rm i} \vec{v}_{\rm i} - n_{\rm e} \vec{v}_{e} ),
\label{current}
\end{eqnarray}
where the coupling of the electromagnetic field to the quintesssence 
field $\sigma$ can be derived from the action
\begin{equation} 
S_{\sigma} = \int d^{4} x \sqrt{-g} \biggl[ \frac{1}{2}g^{\alpha\beta} 
\partial_{\alpha}\sigma \partial_{\beta} \sigma - W(\sigma) + 
\frac{\beta}{4 M} F_{\alpha\beta} \tilde{F}^{\alpha\beta} \biggr],
\label{action}
\end{equation}
where $F_{\alpha\beta}$ and $ \tilde{F}^{\alpha\beta}$ are the Maxwell 
field strength and its dual; $g_{\alpha\beta}$ is the metric tensor. 
In the limit $\sigma \to 0$, 
Eqs. (\ref{div1})--(\ref{current}) are the 
usual curved-space two-fluid equations \cite{f3}.
The evolution of the quintessence field will obey the 
homogeneous equation 
\begin{equation}
\ddot{\sigma} + 3 H \dot{\sigma} + \frac{\partial W}{\partial\sigma} =0,
\label{sigma1}
\end{equation}
where the overdot denotes a derivation with respect to the cosmic 
time coordinate whose differential is related to the conformal 
time coordinate as $ dt = a(\eta) d\eta$; it will be assumed that 
$2 W(\sigma) = m^2\sigma^2$.
From Eqs. (\ref{div1})--(\ref{current}), 
the presence of the quintessence field 
introduces a further frequency scale into the problem, namely 
$\omega_{\sigma} = c (\beta/M) \sigma' $. If the quintessence field 
dominates today (or between redshifts $0$ and $3$), 
$\sigma(t_0) \sim \Lambda^2/m$ 
where, typically, $ \Lambda\sim 10^{-3}\,\, {\rm eV}$, and 
$ m \sim 10^{-33} \,\,{\rm eV}$. The value of $\dot{\sigma}$ can be estimated, 
today, from Eq. (\ref{sigma1}) and it is $ \dot{\sigma}(t_0) \simeq m^2/H_0$.
Hence, recalling that prior to quintessential dominance, $\dot{\sigma}$ scales 
as $a^{-3}$, from the previous expressions  
\begin{equation}
\dot{\sigma}_{\rm dec} \simeq \biggl( \frac{H_{\rm dec}}{H_{0}}\biggr)^2 
\frac{m \Lambda^2}{H_{0}},\,\,\,\,\,\,\,\,\,\,\,
\omega_{\sigma} = \beta \biggl( \frac{M}{M_{\rm P}}\biggr) \times 10^{-6} \,\,
{\rm Hz},
\label{omegasigma}
\end{equation}
where $H_{\rm dec} \sim 10^{6}\,\, H_{0}$; the values of $\Lambda$ 
and $m$ are the ones discussed above with $M\simeq M_{\rm P}$. The value of 
$\beta$ is rather uncertain and a conservative limit from 
radio-astronomical analyses would imply $ \beta \laq 10^{-3}$  for
$0 \laq \,\, z\,\,\laq 1$ \cite{q2}. 
Notice that the action (\ref{action}) can also 
be relevant in a class of baryogenesis models \cite{q4} 
(see also \cite{xinmin}). Similar actions also arise in the framework
of theories with varying coupling constants (see for instance \cite{ralf1}).

Consider the combined effect of the homogeneous 
quintessence field (i.e. $ \vec{\nabla} \sigma =0$) and of a background 
magnetic field on the spectrum of plasma excitations.
Equations (\ref{ne}) and (\ref{np}) together with Equations (\ref{ve})
and (\ref{vi}) 
and (\ref{div1})--(\ref{current}) can then be  linearized in the 
presence of the weak background magnetic field $B_{0}$, i.e. 
\begin{eqnarray}
&& n_{\rm e,\,i}(\eta, \vec{x}) = n_{0} + \delta n_{\rm e,\,i}( \eta,\vec{x}),
\,\,\,\,\,\,\,\,\,\,\, \vec{B}(\eta,\vec{x}) = \vec{B}_{0} + \delta \vec{B}(\eta,\vec{x}),
\nonumber\\
&& \vec{v}_{\rm e,\,i}(\eta,\vec{x}) = 
\delta \vec{v}_{\rm e,\,i} (\eta,\vec{x}),\,\,\,\,\,\,\,\,\,\,\,\,\,\,\,\,\,\,\,\,\,\
\vec{E}(\eta,\vec{x})= \delta \vec{E}(\eta,\vec{x}).
\label{fluct}
\end{eqnarray}
Using Eq. (\ref{fluct}), the system of 
Eqs. (\ref{ne})--(\ref{vi}) and Eqs. (\ref{div1})--(\ref{current}) 
can be written  as
\begin{eqnarray}
&& \delta n_{\rm e}' + n_{0} \vec{\nabla} \cdot \delta \vec{v}_{\rm e} =0,
\,\,\,\,\,\,\,\,\,\, \delta n_{\rm i}' + n_{0} \vec{\nabla} \cdot \delta \vec{v}_{\rm i} =0,
\label{deltanep}\\
&& \delta\vec{v}_{\rm e}' + {\cal H} \delta \vec{v}_{\rm e} = 
- \frac{e}{m_{\rm e}} \biggl[ \delta \vec{E} 
+ \frac{\delta\vec{v_{\rm e}}}{c}  \times \vec{B}_{0}\biggr],
\,\,\,\,\,\,\delta\vec{v}_{\rm i}' + {\cal H} \delta \vec{v}_{\rm i} = \frac{e}{m_{\rm i}} \biggl[ \delta \vec{E} 
+ \frac{\delta\vec{v_{\rm i}}}{c}  \times \vec{B}_{0}\biggr],
\label{deltavep}\\
&&\vec{\nabla} \times \delta \vec{E} = - \frac{1}{c} \delta \vec{B}',
\,\,\,\,\,\,\,\, \vec{\nabla} \cdot \delta \vec{E} = 4 \pi e ( \delta n_{\rm i}
 - \delta n_{\rm e}),
\label{deltadiv}\\
&& \vec{\nabla}\times \delta \vec{B} = \frac{1}{c} \delta \vec{E}'  - 
\frac{\beta}{M} \sigma' \delta \vec{B}
 + \frac{4\pi\,e\,n_{0}}{e} ( \delta \vec{v}_{\rm i} -
 \delta \vec{v}_{\rm e}).
\label{deltacurl}
\end{eqnarray}
From Eqs. (\ref{deltanep})--(\ref{deltadiv}) the relevant dispersion 
relations and the 
associated refraction indices can be obtained by treating 
separately the motions 
parallel and perpendicular to  the magnetic field direction.
Defining the current direction parallel to the magnetic field  as
$\vec{j}_{\parallel}= n_{0} e ( \delta \vec{v}_{{\rm i},\,\parallel} 
- \delta \vec{v}_{{\rm e},\,\parallel})$, 
Eqs. (\ref{deltavep}) imply 
\begin{equation}
\vec{j}_{\parallel}' + {\cal H} \vec{j}_{\parallel} 
=\frac{1}{4\pi} ( \omega_{\rm p,\,i}^2 + \omega_{\rm e,\,i}^2) 
\,\,\delta \vec{E}_{\parallel}.  
\label{jpar}
\end{equation}
Since the variation of the geometry is slow in  with respect to the typical 
frequencies of plasma oscillations,
 the following adiabatic expansions can be used:  
\begin{equation} 
\vec{j}_{\parallel}(\eta, \vec{x}) =\vec{j}_{\parallel,\omega}(\vec{x}) e^{- i \int^{\eta} d\eta' \omega(\eta')},\,\,\,\,\,\,\,\,\,\,\,\,\,
 \delta \vec{E}_{\parallel}(\eta, \vec{x}) =  \delta \vec{E}_{\parallel,\omega}( \vec{x})
e^{- i \int^{\eta} d\eta' \omega(\eta')}.
\end{equation}
Thus, defining $\alpha = i {\cal H}/\omega\ll 1$, Eq. (\ref{jpar}) implies that
\begin{equation}
\vec{j}_{\parallel, \omega} = \frac{i}{4\pi} \frac{\omega_{\rm p,\,i}^2 
+ \omega_{\rm p,\,e}^2}{\omega ( 1 + \alpha)} \delta \vec{E}_{\parallel,\omega}.
\label{jpar2}
\end{equation}
Inserting Eq. (\ref{jpar2}) into the parallel component of 
Eq. (\ref{deltacurl}), the following equation can be obtained:
\begin{equation}
(\vec{\nabla}\times \delta \vec{B}_{\omega})_{\parallel} = - i\frac{\omega}{c} \epsilon_{\parallel}(\omega,\alpha) \delta \vec{E}_{\parallel,
\omega} - \frac{\beta}{M}
\sigma' \delta \vec{B}_{\parallel,\omega},
\end{equation}
where  the parallel dielectric constant is 
\begin{equation}
\epsilon_{\parallel}(\omega,\alpha) = 1 - \frac{ \omega_{\rm p,\,i}^2}{\omega^2 (1 +\alpha)} - \frac{ \omega_{\rm p,\,e}^2}{\omega^2 (1 +\alpha)}.
\label{epspar}
\end{equation}
Notice that only the leading curved-space correction has been kept.

With a similar procedure, also the motion in the plane 
 orthogonal to the magnetic field 
direction can be  solved  and  the evolution equations of the electric and 
magnetic fluctuations can then  be written, in compact notation, as
\begin{eqnarray}
&& \vec{\nabla} \times 
\delta \vec{E}_{\omega} = i \frac{\omega}{c} \delta \vec{B},
\label{red1}\\
&& \vec{\nabla} \times \delta \vec{B}_{\omega} = 
- i \frac{\omega}{c} \overline{\epsilon}(\alpha,\omega) \delta 
\vec{E}_{\omega} - \frac{\beta}{M} \sigma' \delta \vec{B}_{\omega},
\label{red2}
\end{eqnarray}
where $\delta \vec{E}_{\omega}$ and $\delta \vec{B}_{\omega}$ have to 
be understood as column matrices containing, in each row, the components 
of the electric and magnetic fields in each of the three spatial 
directions, while $\overline{\epsilon}(\omega,\alpha)$ 
is a $3\times 3$ matrix given by 
\begin{equation}
\overline{\epsilon}(\omega,\alpha) 
= \left(\matrix{\epsilon_{1}(\omega,\alpha)
& i\epsilon_{2}(\omega,\alpha) & 0&\cr
-i \epsilon_{2}(\omega,\alpha) & \epsilon_{1}(\omega,\alpha) &0&\cr
0&0&\epsilon_{\parallel}(\omega,\alpha) }\right),
\label{epstens}
\end{equation}
where $\epsilon_{\parallel}(\omega,\alpha)$ is defined by Eq. (\ref{epspar});
 $ \epsilon_{1,2}(\omega,\alpha)$ are instead 
\begin{eqnarray}
&& \epsilon_{1}(\omega,\alpha) = 1 - \frac{\omega^2_{\rm p\, i} (\alpha + 1) }{\omega^2 (\alpha+ 1)^2 - \omega_{\rm B\, i}^2} -
\frac{\omega^2_{\rm p\, e} (\alpha + 1) }{\omega^2 (\alpha+ 1)^2 - \omega_{\rm B\, e}^2},
\label{eps1}\\
&& \epsilon_{2}(\omega,\alpha) = \frac{\omega_{\rm B\, e}}{\omega} \frac{\omega^2_{\rm p\, e } }{\omega^2 (\alpha+ 1)^2 - \omega^2_{\rm B\, e}} 
-  \frac{\omega_{\rm B\, i}}{\omega} \frac{\omega^2_{\rm p\, i } }{\omega^2 (\alpha+ 1)^2 - \omega^2_{\rm B\, i}} .
\label{eps2}
\end{eqnarray}
The coordinate system can be fixed by setting $k_{x}=0$ and 
$k_{y} = k \sin{\theta}$, $k_{z} = k \cos{\theta}$ with $\vec{B}_{0}$ 
oriented along the $\hat{z}$ direction.
Since Eqs. (\ref{red1}) and (\ref{red2}) imply
\begin{equation}
\vec{\nabla} \times \vec{\nabla} \times \delta \vec{B}_{\omega} = 
\frac{\omega^2}{c^2} \overline{\epsilon}(\omega,\alpha) \delta 
\vec{B}_{\omega} - \frac{\beta}{M} \sigma' \vec{\nabla} \times \delta 
\vec{B}_{\omega},
\label{red3}
\end{equation}
the Fourier transform of Eq. (\ref{red3}), in the coordinate system 
selected previously, leads  to the generalized 
Appleton--Hartree equation:
\begin{equation}
{\cal A}\,\,\, \delta\vec{B}_{\vec{k},\omega}
= \left(\matrix{\bigl[1 - \frac{\epsilon_{1}}{n^2}\bigr]
& - i\bigl[\frac{\epsilon_{2}}{n^2} 
+ \frac{\omega_{\sigma}}{n \omega} c(\theta) \bigr] & i
 \frac{\omega_{\sigma}}{n \omega} s(\theta) &\cr
i\bigl[\frac{\epsilon_{2}}{n^2} 
+ \frac{\omega_{\sigma}}{n \omega} c(\theta) \bigr] 
& \bigl[ c^2(\theta) - \frac{\epsilon_1}{n^2} \bigr] 
& - s(\theta) c(\theta)&\cr
- i \frac{\omega_{\sigma}}{n \omega} s(\theta) & - s(\theta)c(\theta)
&\bigl[ s^2(\theta) -\frac{\epsilon_{\parallel}(\omega,\alpha)}{n^2}\bigr] }\right) \left(\matrix{\delta B_{k,\omega,x}\cr
\delta B_{k,\omega,y}\cr
\delta B_{k,\omega,z}}\right)=0,
\label{Amatrix}
\end{equation}
 where the refraction 
index $n = c/v$ has been introduced so as to eliminate 
the comoving momentum, in such a way that $k = \omega/v = n \omega/c$; 
we have written $c(\theta) = \cos{\theta}$ and $s(\theta) = \sin{\theta}$.
From Eq. (\ref{Amatrix}), $ {\cal A}^{\dagger} = {\cal A}$, where 
the dagger denotes the transposed and complex conjugate of a given matrix.

The non-trivial solutions of the system of algebraic (homogeneous) 
equations given by 
formula (\ref{Amatrix}) comes from setting the determinant of the 
coefficients equal to zero, i.e. ${\rm det} {\cal A} =0$. 
After some algebra it is found that the determinant vanishes if 
\begin{eqnarray}
&&s^2(\theta)\biggl\{ \biggl( \frac{1}{\epsilon_{\parallel}} - \frac{1}{n^2} 
\biggr) \biggl[ \frac{1}{n^2} - \frac{1}{2} \biggl( \frac{1}{\epsilon_{\rm L}}
+ \frac{1}{\epsilon_{\rm R}}\biggr) \biggr] + 
\frac{\omega_{\sigma}^2}{2 n^2\omega^2}\biggl(\frac{1}{\epsilon_{\rm L}}
+ \frac{1}{\epsilon_{\rm R}} \biggr) 
\nonumber\\
&&-
c^2(\theta) \biggl[\biggl( \frac{1}{n^2} - \frac{1}{\epsilon_{\rm L}}
\biggr)\biggl( \frac{1}{n^2} - \frac{1}{\epsilon_{\rm R}}\biggr) - 
\frac{\omega_{\sigma}^2}{n^2 \omega^2\epsilon_{\rm R} \epsilon_{\rm R}}\biggr]
+ \frac{\omega_{\sigma}}{n^3 \omega}\biggl( \frac{1}{\epsilon_{\rm L}} - 
\frac{1}{\epsilon_{\rm R}}\biggr) c(\theta) =0,
\label{AH}
\end{eqnarray}
where the right-handed and left-handed dielectric constants have been 
defined as 
\begin{eqnarray}
&& \epsilon_{\rm R} = \epsilon_{1} + \epsilon_{2} = 1 - \frac{ \omega^2_{\rm p\,i}}{\omega[ \omega (\alpha + 1) - \omega_{\rm B\,i}]} 
-\frac{ \omega^2_{\rm p\,e}}{\omega[ \omega (\alpha + 1) 
+ \omega_{\rm B\,e}]},
\label{epsR}\\
&& \epsilon_{\rm L} = \epsilon_{1} - \epsilon_{2} = 1 -  \frac{ \omega^2_{\rm p\,e}}{\omega[ \omega (\alpha + 1) -\omega_{\rm B\,e}]}
- \frac{ \omega^2_{\rm p\,i}}{\omega[ \omega (\alpha + 1) 
+ \omega_{\rm B\,i}]}.
\label{epsL}
\end{eqnarray}
If $\omega_{\sigma} =0$, then Eq. (\ref{AH}) reduces exactly to the 
Appleton--Hartree equation known from two-fluid plasma theory \cite{pl1}, with 
the minor difference that the leading dependence upon the background geometry 
appears in $\epsilon_{\rm R,L}$  through the function $\alpha$ . 
The dispersion relations for a wave propagating parallel and 
perpendicular to the magnetic field 
direction can be obtained by setting, respectively,  
$ \theta =0$ and $\theta = \pi/2$ in Eq. (\ref{AH}). Consequently, the 
relevant equations determining the refraction index are, in this case 
\begin{eqnarray} 
&& (n^2 - \epsilon_{\rm R})(n^2 -\epsilon_{\rm L})=0,\,\,\,\,\,\,
\,\,\,\,\,\,\,\,\,\,\,\,\,\,\,\,\,\,\,\,\,\,\,\,\,\,\,\,\theta=0,
\label{std0}\\
&& (n^2 - \epsilon_{\parallel}) [ n^2 (\epsilon_{\rm L} + \epsilon_{\rm R})
- 2 \epsilon_{\rm L} \epsilon_{\rm R} ] =0,
\,\,\,\,\, \theta = \frac{\pi}{2}.
\label{stdpi2}
\end{eqnarray}
Equation (\ref{std0}) gives the usual dispersion relations for the two circular 
polarizations of the electromagnetic wave, i.e. $n^2 = \epsilon_{\rm R}$ 
and $n^2 = \epsilon_{\rm L}$, while Eq. (\ref{stdpi2}) 
gives those for the ``ordinary'' 
(i.e. $n^2 = \epsilon_{\parallel}$) and ``extraordinary'' (i.e. $n^2 = 
2 \epsilon_{\rm R} \epsilon_{\rm L}/(\epsilon_{\rm R} + \epsilon_{\rm L})$)
plasma waves \cite{pl1,pl2}. 

In the presence of a dynamical quintessence field weakly coupled 
to electromagnetism,  $\omega_{\sigma} \neq 0$ and the resulting 
dispersion relations are, according to Eq. (\ref{AH}): 
\begin{eqnarray}
&& \biggl( n^2 -  \frac{\omega_{\sigma}}{\omega}n - \epsilon_{\rm R}\biggr) 
\biggl(n^2 + \frac{\omega_{\sigma}}{\omega} n - \epsilon_{\rm L}\biggr) =0,
\,\,\,\,\,\,\,\,\,\,\,\,\,\,\,\,\,\,\,\,\,\,\,\,\,\,\,\,\theta =0,
\label{nsd0}\\
&& n^4 - \biggl[ 2 \frac{\epsilon_{\rm L}\epsilon_{\rm R}}{\epsilon_{\rm L} 
+ \epsilon_{\rm R}} + \epsilon_{\parallel} 
+ \biggl(\frac{\omega_{\sigma}}{\omega}
\biggr)^2\biggr] n^2 + 2 \frac{\epsilon_{\parallel} \epsilon_{\rm L} 
\epsilon_{\rm R}}{ \epsilon_{\rm R} +\epsilon_{\rm L}} =0,
\,\,\,\,\,\theta = \frac{\pi}{2}.
\label{nsdpi2}
\end{eqnarray}

From Eq. (\ref{nsd0}) the generalized Faraday rotation experienced 
by the linearly polarized CMB travelling parallel to the magnetic 
field direction can be obtained as 
\begin{equation}
\Delta\Phi = \frac{\omega}{2c} \biggl[ \frac{\omega_{\sigma}}{\omega}
+ \sqrt{\frac{1}{2}  \biggl(\frac{\omega_{\sigma}}{\omega}\biggr)^2 
+ \epsilon_{\rm R}} -\sqrt{\frac{1}{2}  \biggl(\frac{\omega_{\sigma}}{\omega}\biggr)^2 + \epsilon_{\rm L}} \biggr] \Delta {\rm L} ,
\end{equation}
where $\omega \simeq \omega_{\rm CMB}$ at decoupling; $\Delta {\rm L}$ is 
the distance travelled by the signal in the direction parallel to the 
magnetic field direction. 
It is interesting to compare the contribution of the terms depending 
upon $\omega_{\sigma}$ and those depending upon the background magnetic 
field intensity, i.e. the terms appearing in the squared brackets.

According to Eqs. (\ref{plei}),(\ref{Bei}) and (\ref{omegasigma}) 
the leading contribution to the generalized Faraday rotation 
arises as the sum of two dimensionless ratios, i.e. 
\begin{equation}
\biggl(\frac{\omega_{\sigma}}{\omega_{\rm CMB}}\biggr) + 
\biggl(\frac{\omega_{\rm Be}}{\omega_{\rm CMB}}\biggr) 
\biggl(\frac{\omega_{\rm pe}}{\omega_{\rm CMB}}\biggr)^2.
\label{cont}
\end{equation}
Using Eqs. (\ref{plei}),(\ref{Bei}) and (\ref{omegasigma}),
it can be verified directly that the two contributions appearing in Eq. (\ref{cont}) 
are of the same order, provided 
\begin{equation}
\biggl(\frac{\beta}{10^{-3}}\biggr) \biggl(\frac{M}{M_{\rm P}}\biggr) 
\simeq \biggl(\frac{B_{0}}{10^{-4}\,\,{\rm G}}\biggr) \biggl(\frac{n_{\rm e}}{10^{-3}\,\,\,{\rm cm}^3}\biggr),
\label{comp}
\end{equation}
where in the left-hand-side we computed the quintessential contribution,
 at the right-hand-side the contribution of the magnetic field has 
been reported. The values of $\beta$ and $M$ are, 
as previously stressed, conservative estimates, compatible with present bounds
\cite{q2}.
 For the typical (allowed) parameters reported in Eq. 
(\ref{comp}), the two contributions then are, as anticipated, of the same order.
For the explicit value of $\omega_{\rm CMB}$ the numerical value 
$\omega_{\rm CMB} \sim {\rm THz}$ has been used (notice, in fact, that 
the maximum of the CMB is of the order of the ${\rm GHz}$ {\em today}
and of the order of the ${\rm THz}$ around decoupling). Given the possibility
of absence of hierarchy between the two dimensionless ratios, Taylor 
expansions of the obtained disperion relations should be treated with care, 
especially in view of possible experimental applications of these findings.
In a complementary perspective, when analysing the possible 
rotation of the CMB polarization,  it seems then preferable to adopt 
the generalized formulae derived in the present study.

Concerning the obtained results, a few comments are in order.
\begin{itemize}
\item{} the two contributions have a frequency dependence that may allow us  
to disentangle their relative weight since the magnetic contribution 
vanishes for $\omega >\omega_{\rm CMB}$ as $1/\omega^2$;
\item{} for the set of parameters chosen in Eq. (\ref{comp}) 
the typical generalized Faraday {\em rotation measurement} (i.e. 
$\Delta \Phi = {\rm RM}\,\, \lambda_{\rm CMB}^2$, $\lambda$ being the 
wavelength) will be of the order of $200 \,\,{\rm rad}/{\rm m}^2$ 
if we use the 
fact that the optical depth 
for Thompson scattering at decoupling is $\int x_{\rm e} n_{\rm e}a(\eta) 
 d\eta \sim \sigma_{\rm T}^{-1}$ where $\sigma_{\rm T}$ is the Thompson 
cross section;
\item{} situations can be envisaged ($\beta \ll 10^{-3}$) 
where the quintessential contribution
is far smaller than the ``magnetic'' contribution;
\item{} the dispersion relations for radiation propagating orthogonal to 
the magnetic field intensity also lead, in the case $\omega_{\sigma}\neq 0$, 
to a rotation where the magnetized contribution is absent and $\Delta\Phi
\propto \omega_{\sigma} \Delta {\rm L}$;
\item{} in the (generalized) extraordinary dispersion relations 
(see Eq. (\ref{nsdpi2})
 a resonance may
appear, when, $\beta \ll 10^{-3}$ and for $\omega_{\rm r} \simeq 
\sqrt{\omega_{\rm pe}^2 + \omega_{\rm Be}^2}$; however $\omega_{\rm r} 
\ll \omega_{\rm CMB}$;
\end{itemize}

The derivation presented so far refers to the case of cold 
magnetoactive plasmas. The effects arising from the finite temperature of the
 electrons do not modify the leading result obtained in the context of the 
cold plasma theory. The derivation of the dispersion relations 
in the case of a warm plasma can be performed, for instance, within a 
kinetic approach where it can be shown, following the same 
calculation discussed in the 
flat space case \cite{skilling}, 
that the first correction to the leading cold plasma calculation can be
 recast in an effective redefinition of the 
plasma frequency for the electrons, namely
$\omega_{\rm p\,e} \to \frac{4 \pi n_{0} e^2}{m_{e} \gamma}$
where $\gamma = (1 - \langle v^2\rangle)^{-1/2}$  and $\langle v^2 \rangle$ is the thermal 
average of the electron velocity.

\end{document}